\documentclass[journal]{IEEEtran}
\usepackage{cite}

\ifCLASSINFOpdf
   \usepackage[pdftex]{graphicx}
  \graphicspath{{./figures/}}
   \DeclareGraphicsExtensions{.pdf,.jpeg,.png}
\else
\fi
\usepackage[cmex10]{amsmath}
\begin{document}
%
\title{Fast Scatter Artifacts Correction for Cone-Beam CT without System Modification and Repeat Scan}
%
%
%

\author{Wei~Zhao,~Jun Zhu, Luyao Wang
\thanks{W. Zhao~(e-mail: zhaow85@163.com), J. Zhu and LY. Wang are with the Department of Biomedical Engineering, Huazhong University of Science and Technology, Hubei, China.}
}

\maketitle


\begin{abstract}
We provide a fast and accurate scatter artifacts correction algorithm for cone beam CT (CBCT) imaging.
The method starts with an estimation of coarse scatter profile for a set of CBCT images. A total-variation denoising algorithm designed specifically for Poisson signal is then applied to derive the final scatter distribution. Qualitatively and quantitatively evaluations using Monte Carlo (MC) simulations, experimental CBCT phantom data, and \emph{in vivo} human data acquired for a clinical image guided radiation therapy were performed. Results show that the proposed algorithm can significantly reduce scatter artifacts and recover the correct HU within either projection domain or image domain. Further test shows the method is robust with respect to segmentation procedure.
\end{abstract}


%
\IEEEpeerreviewmaketitle

\section{Introduction}
%
%
%
%
\IEEEPARstart{D}{ue} to the increased axial coverage of multi-slice computed tomography (CT) and the introduction of flat-panel detector, the size of X-ray illumination fields has grown dramatically, causing an increase in scatter radiation. For most of cases, artifacts are present when the model used in the reconstruction algorithm is not consistent with the projection data acquisition model. Existing reconstruction algorithms don't model the scatter radiation, so scatter artifacts appear in the CT images. Typical scatter artifacts show as shading, streaks between high contrast objects, reduced contrast resolution and inaccurate Hounsfield Units (HU) numbers. Scatter correction has been extensively studied in the past serval decades but a clinically sensible solution remains illusive. Current scatter correction methods can be roughly classified into five approaches: physical scatter rejection, analytical modeling, Monte Carlo (MC) simulation, primary modulation and scatter measurements~\cite{zhu2009}.

To develop a scatter artifacts correction method that is fast, accurate, and, ideally, has no need for extra scans, extra measurements or extra hardware support or modifications is the goal of this study. We provide a fast and accurate scatter artifacts correction method for cone beam CT (CBCT) imaging. In the proposed method, the difference between the measured raw projection data and a polychromatic reprojection of a segmented image volume is used to generate a coarse estimate of scatter profile. Based on the coarse scatter estimation, a denoising algorithm specifically designed for Poisson signal which employs a total variation regularization term was then applied to yield the final scatter estimation. By simply subtracting the final scatter estimate from the raw projection data, we can obtain the corrected projection data. The corrected image was then reconstructed using the corrected projection data.
MC simulations, experimental phantom data and \emph{in vivo} human data were used to validate and evaluate the proposed scatter correction method.

\section{Material and methods}
The measured raw projection data can be modeled as the summation of the primary and scatter signals. A general scatter artifact correction strategy is to estimate the scattered radiation distribution and then subtract it from the measured raw projection data, i.e.
\begin{equation}
I_{p}(\alpha,\vec{x}) = I(\alpha,\vec{x})-I_{s}(\alpha,\vec{x}),
\end{equation}
where $I_{p}$ is the corrected projection data, namely, the estimated primary projection data, $I$ is the raw data and $I_s$ is the scatter signal. Indices $\alpha $ and $\vec{x}$ stand for projection view angle and detector channel number respectively. For simplicity, we will drop the index $\vec{x}$ in the following discussions.

\subsection{Coarse scatter estimation}
\begin{figure}[t]
    \centering
    \includegraphics[width=3.5in]{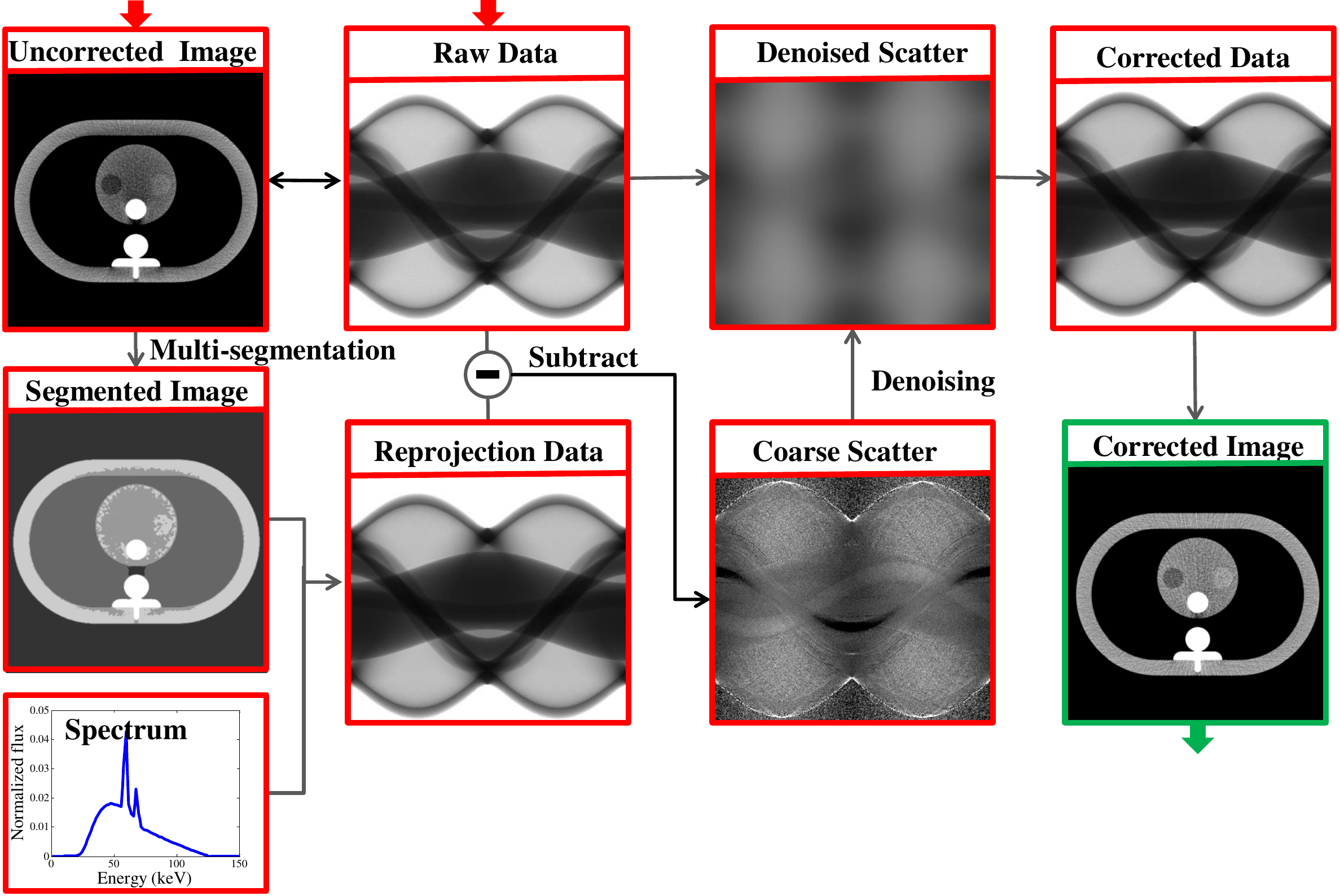}
    \caption{Flowchart of the proposed scatter correction method. The method can starts with either the raw projection data or an scatter contaminated volume as input. After an initial CT image reconstruction, the image is segmented into different components, based on which a polychromatic reprojection is performed with predetermined spectrum. The reprojection data is subtracted from the raw projection data to yield a coarse scatter estimate. The coarse scatter is then applied to a denoising algorithm which was specifically designed for Poisson signal to yield the final scatter estimate. The corrected projection data is obtained by subtracting the final scatter from the raw projection data and the corrected CT image can be generated using a standard filtered backprojection reconstruction algorithm. }
    \label{fig:f1}
    \vspace{-2mm}
\end{figure}

The proposed method, as shown in Fig.~\ref{fig:f1}, starts with an image segmentation of the scatter artifacts contaminated CBCT image volume, followed by a polychromatic reprojection of the segmented image volume using a given x-ray spectrum  and known data acquisition geometry. The reprojected data is then subtracted from the measured raw projection data at each given view angle to generate a coarse estimate of the necessary scatter profile used for denoising algorithm.

Based on the segmented image volume, the primary signal can be modeled as follows:
\begin{equation}\label{equ:reporjection}
\hat{I}_{p}=N\int_{0}^{E_{max}}\mathrm{d}E\,\Omega(E) \, \eta(E)\,\mathrm{exp}\left[-\int_{0}^{l}\mu(E,s)\mathrm{d}s\right],
\end{equation}
where $N$ is the total number of photons, $\Omega(E)$ is the polychromatic X-ray spectrum, $\eta(E)$ is the energy-dependent efficiency of the detector, which is simply considered as proportional to photon energy $E$ because most CT scanners use energy-integrating detectors. $E_{max}$ is the maximum photon energy of the spectrum. $\mu(E,s)$ is the energy-dependent linear attenuation coefficient and $l$ is the propagation path length for each ray, and can be calculated using a GPU-based ray-tracing algorithm~\cite{pratx2011,jia2014}.
By using these notations, the flood field $I_{0}$ can be modeled as follows:
\begin{equation}\label{equ:floodfield}
I_{0}=N\int_{0}^{E_{max}} \mathrm{d}E\,\Omega(E)\,\eta(E).
\end{equation}
After replacing $N$ with $I_{0}$ and substituting it into~(\ref{equ:reporjection}), the result describes the estimated primary signal with flood field $I_{0}$:
\begin{equation}\label{equ:primary}
\hat{I}_{p}=\frac{I_{0}\int_{0}^{E_{max}}\mathrm{d}E\,\Omega(E)\,\eta(E)\,\mathrm{exp}\left[-\int_{0}^{l}\mu(E,s)\mathrm{d}s\right]}{\int_{0}^{E_{max}}\mathrm{d}E\,\Omega(E)\,\eta(E)}.
\end{equation}
Subtracting the above estimated primary signal from the measured total projection data $I$,  the coarse scatter signal $\hat{I}_s$ can be estimated as follows:
\begin{equation}\label{equ:coarseScatter}
\hat{I}_{s}=I-\frac{I_{0}\int_{0}^{E_{max}}\mathrm{d}E\,\Omega(E)\,\eta(E)\,\mathrm{exp}\left[-\int_{0}^{l}\mu(E,s)\mathrm{d}s\right]}{\int_{0}^{E_{max}}\mathrm{d}E\,\Omega(E)\,\eta(E)}.
\end{equation}
Note that the segmented image volume is generated from a scatter contaminated reconstruction and the polychromatic x-ray spectrum used in the reprojection can be recovered using an indirect transmission measurement-based spectrum estimation method~\cite{zhao2015}.  

\vspace{-2mm}
\subsection{Denoising the coarse scatter}
Since the coarse scatter estimate $\hat{I}_s$ is highly dependent on the segmentation procedure, it may yield inaccurate results, especially for low contrast objects that have similar attenuation properties as the background and for the edges of two neighboring materials.  To compensate for the inaccuracy caused by the inaccurate segmentation, instead of regularizing the coarse scatter using a convolution-based scatter model~\cite{zhao2014}, in this study, we directly denoise the coarse scatter estimate
using a statistical-based denoising algorithm.

As we know, scatter radiation distribution is typically predominantly of low frequency content not only in spatial domain but also in temporal domain. 
A denoising technique specifically designed for Poisson signal was applied to $\hat{I}_s$ to yield a smooth scatter distribution~\cite{jia2012}.  The denoising algorithm which is based on the total-variation regularization is aimed to solve the following optimization problem,
\begin{equation}\label{equ:energyfunction}
I_s=\mathrm{argmin}_{I_s}\;\int \mathrm{d}\vec{x}(I_s-\hat{I}_s \mathrm{log}I_s)+\frac{\beta}{2}\int d\vec{x}|\nabla I_s|^{2}.
\end{equation}
The first term of ~(\ref{equ:energyfunction}) is a data-fidelity term that is designed specifically for Poisson statistical signal and keep $I_s$ close to the data $\hat{I}_s$, while the second term is a total-variation regularization term to keep the solution $I_s$ smooth, namely, being dominated with low frequency content. $\beta$ is a constant to determine the relative importance of the two terms. This objective function is convex and can be solved using an variational approach~\cite{jia2012}.

%

It has to note that the proposed method can also be implemented in image-domain. For those cases when we only have images or volume, we need to estimate scatter artifact error images which can be added directly to the uncorrected images. To achieve this goal, a scatter projection error $\Delta p_s$ which can be added linearly in the logarithmic raw-data domain can be calculated. Because the tomographic reconstruction is a linear process and the order of summation and backprojection operations are interchangeable.

\vspace{-2mm}
\subsection{Monte Carlo simulations}
To validate the proposed algorithm, an anthropomorphic\footnote{http://www.qrm.de/content/pdf/QRM-Thorax.pdf.} thorax phantom and an abdomen phantom were used to generate MC simulation data with a Geant4-based MC simulation package GATE~\cite{Jan2011}. To quantitatively and qualitatively investigate the effect of scatter artifacts reduction, both of the phantoms include a water insert at the central area and the water insert has three small low contrast inserts (adipose, breast and liver) and a bone insert. 

Primary projection data, scatter only projection data, and primary plus scatter projection data were extracted independently. 
For a faster simulation, a parallel geometry and a plane X-ray source (2D $320\times120$~mm$^{2}$ rectangle source) were used in the MC simulation. The distance from the source to the center of rotation is 750 mm and the distance from the detector to center of rotation is 450 mm. A circular scan was simulated and a total of 360 projections per rotation are acquired over an angular range of $360^{\circ}$.
The polychromatic X-ray source spectrum is 125 kVp and it is generated using Spektr software~\cite{siewerdsen2004b} with 5 mm aluminum filtration. For each of the simulations, a total number of $3\times10^{10}$ photons were emitted.


In a first experiment, we investigate scatter correction for the anthropomorphic thorax phantom and abdomen phantom in both projection domain and image domain.
The primary image was used as reference images and the total images were applied to the proposed scatter correction algorithm.



To test the robustness of method with respect to the segmentation, two components (lung and tissue), three components (lung, tissue and bone) and four components (lung, water, tissue and bone) segmentation were performed based on the uncorrected thorax phantom images respectively. 

%
\vspace{-2mm}
\subsection{Physical phantom experiments and patient study}
The proposed algorithm was also evaluated using experimental data of a Catphan500 phantom (The Phantom Laboratory, Salem, NY) scanned using a CBCT on-board imager (Varian 2100EX System, Varian Medical Systems, Palo Alto, CA).
A total of 678 projections were evenly acquired in a 360 degree rotation with $2\times2$ binning. 
Both wide collimation and narrow collimation modes were applied with the same scan parameters.  



A set of \emph{in vivo} pelvis data was also used to evaluate the proposed method. This data was acquired in half-fan mode using state-of-the-art of the kV CBCT imaging system of the Varian TrueBeam system(Varian Medical Systems, Palo Alto, USA). The tube potential was set to 125 kVp and 678 projections were acquired on a flat-panel detector (Varian 4030CB imager) with 1024 columns and 768 rows. 

\begin{figure}
    \centering
    \includegraphics[width=0.5\textwidth]{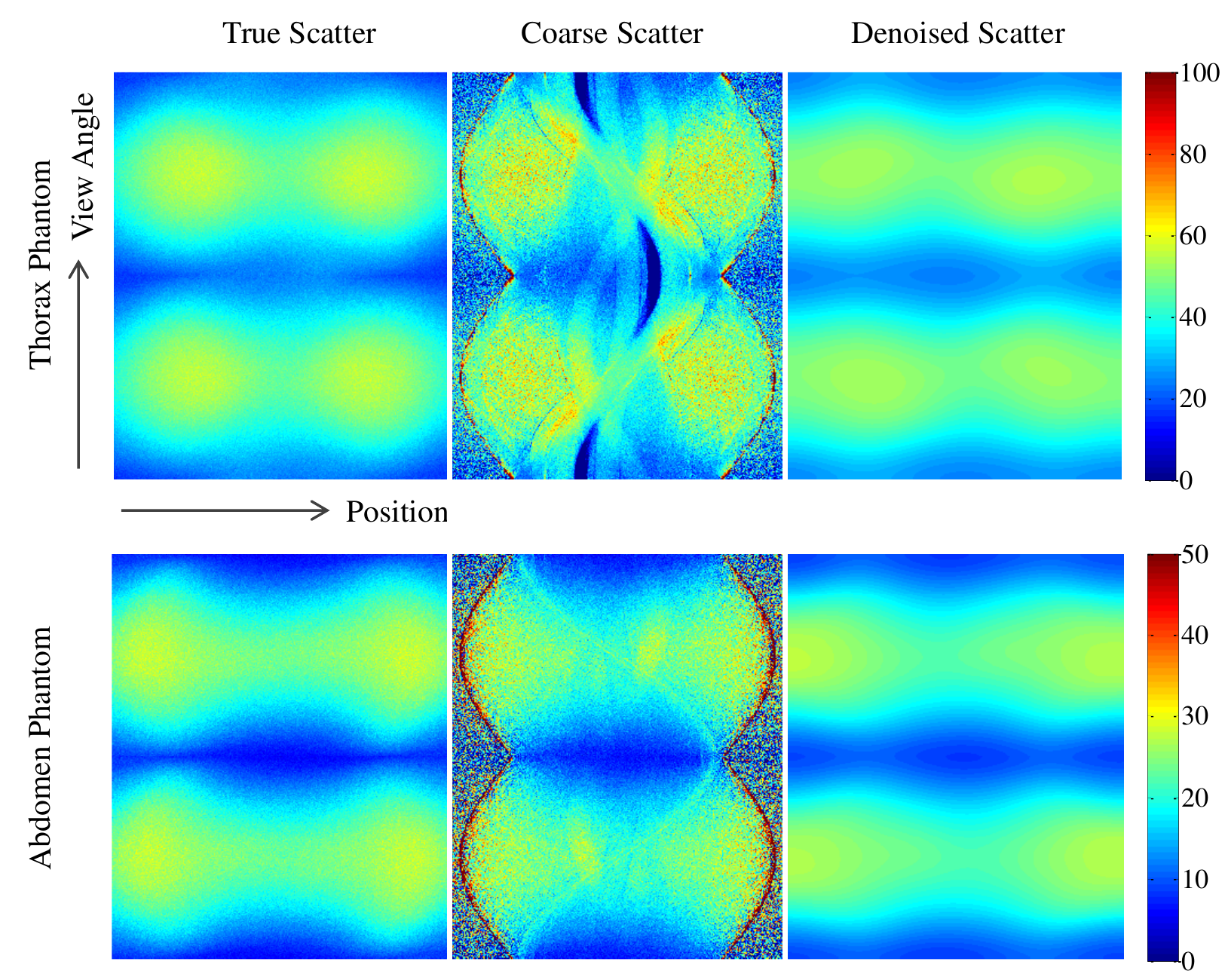}
    \caption{Sinograms of true scatter, coarse scatter, and denoised scatter for an anthropomorphic thorax phantom and an abdomen phantom. The coarse scatter which is the difference of the raw projection data and the polychromatic reprojection data have global scatter profiles but also contain errors which were partially generated from inaccurate segmentation. The denoised coarse scatter smooth these errors and fit the true scatter profiles well.}
    \label{fig:scatterProfile}
    \vspace{-3mm}
\end{figure}

\section{RESULTS} \label{sec:sections}

\subsection{Monte Carlo simulations}



\begin{figure}[t]
    \centering
    \includegraphics[width=3.5in]{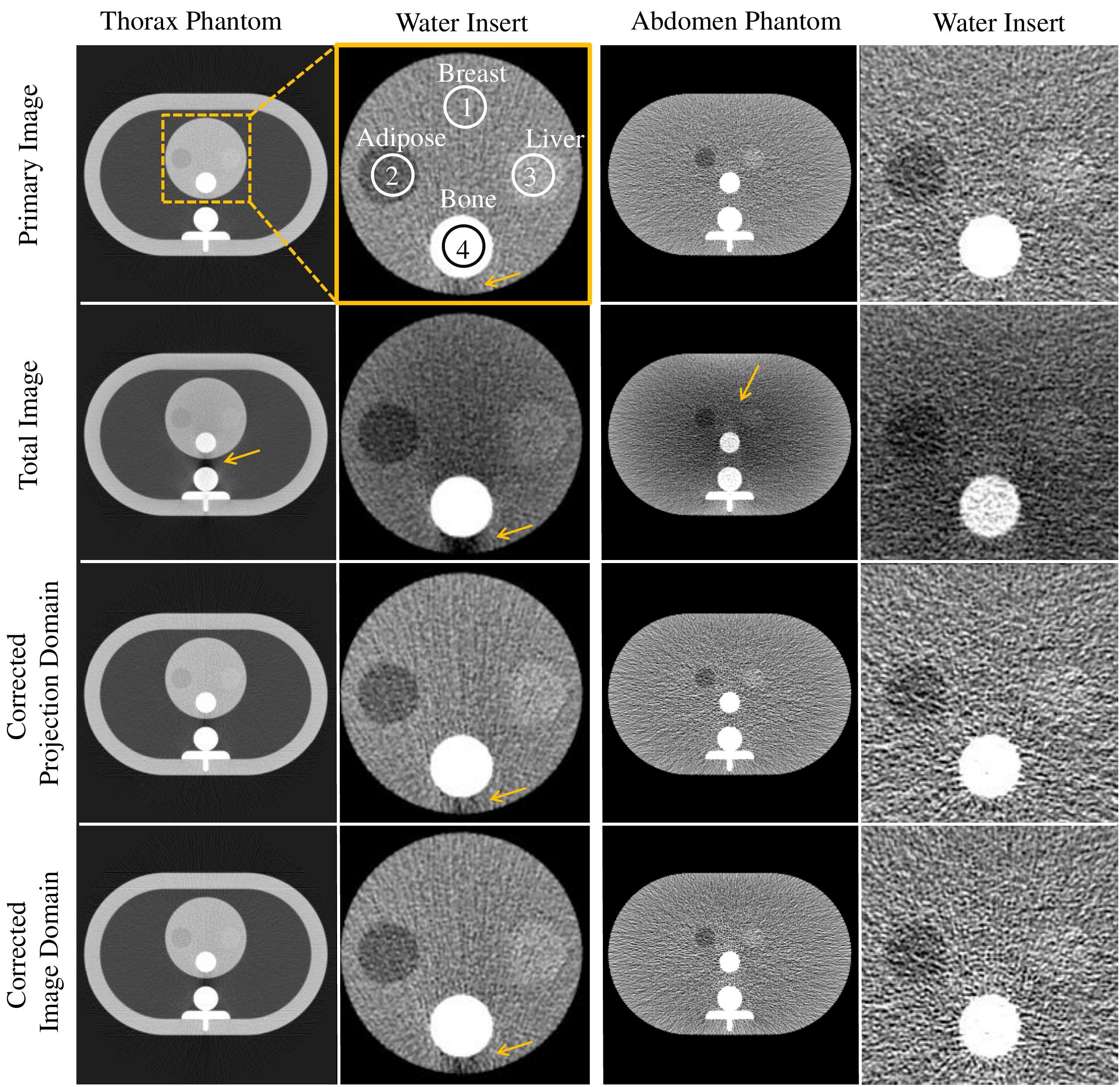}
    \caption{Results of the scatter correction for the MC simulation data of the thorax phantom and the abdomen phantom within both projection domain and image domain. (First row) Results for the primary signal and they can be regarded as ground truth. (Second row) Results for the primary plus scatter signal (total signal) and scatter induced shading artifacts and streaks are clearly seen. (Third row) Results for the scatter corrected images within projection data domain. (Fourth row) Results for the scatter corrected images within image domain. Shading artifacts and streaks are significantly reduced after correction within both projection domain and image domain. Display window: [-1200HU, 500HU] for thorax phantom images and [-300HU, 300HU] for the water insert images and the abdomen phantom.}
    \label{fig:domains}
\end{figure}

\begin{figure}[t]
    \centering
    \vspace{-3mm}
    \includegraphics[width=3.5in]{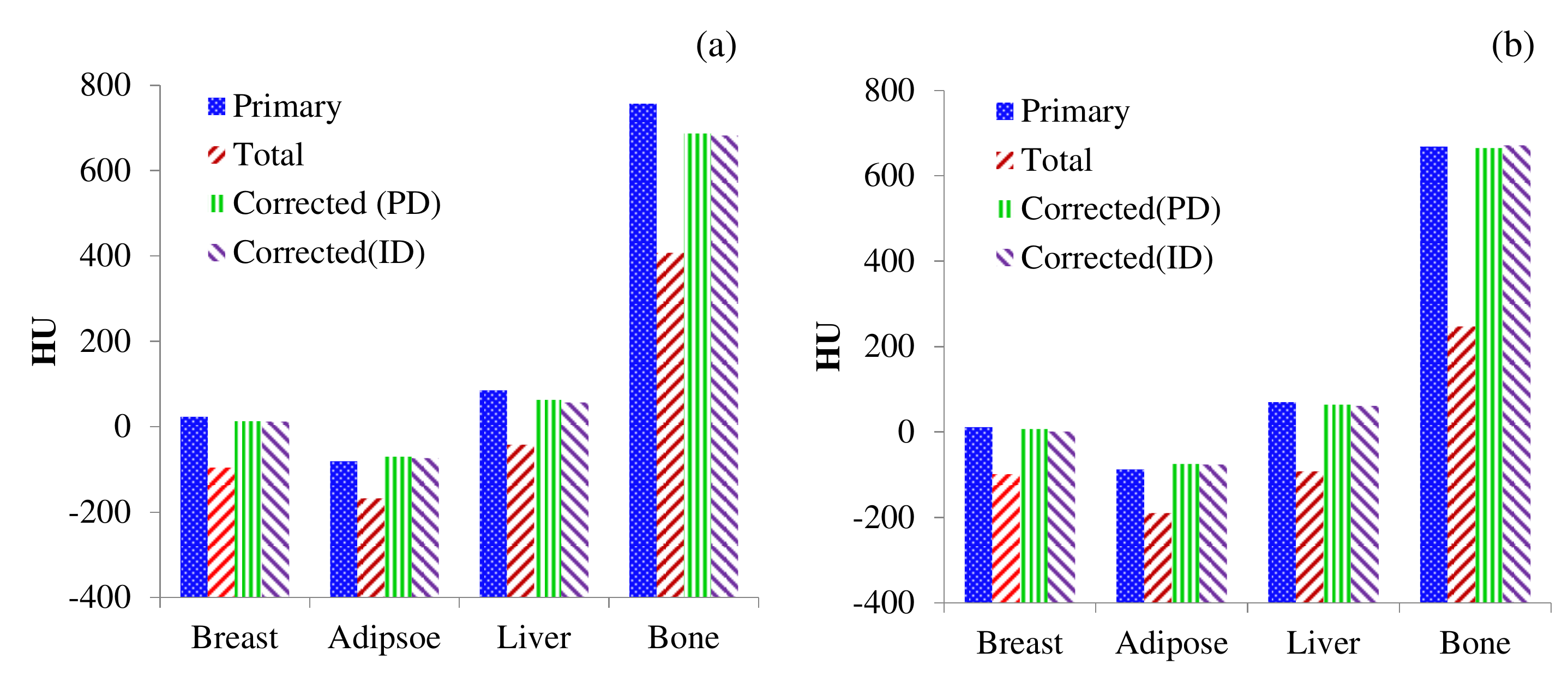}
    \caption{Results of reconstruction values in HU numbers at different ROIs of the water insert of (a) the thorax phantom and (b) the abdomen phantom before and after correction within both projection domain and image domain. The HU numbers are successfully recovered after scatter correction using the proposed method. PD and ID stand for projection domain and image domain, respectively. }
    \label{fig:domainsHU}
    \vspace{-5mm}
\end{figure}
Fig.~\ref{fig:scatterProfile} shows the sinograms of the MC reference scatter, coarse scatter and the denoised scatter for the anthropomorphic thorax phantom and the abdomen phantom.
Fig.~\ref{fig:domains} shows the results of the scatter correction using the proposed method for the MC simulation data of the thorax phantom and the abdomen phantom within both projection domain and image domain. The primary images were reconstructed using primary projections and they were regarded as scatter-free images.
Fig.~\ref{fig:domainsHU} depicts the HU numbers of breast, adipose, liver and bone inserts (shown in Fig.~\ref{fig:domains}) of primary images, total images and scatter corrected images. 

\begin{figure}
    \centering
    \includegraphics[width=3.5in]{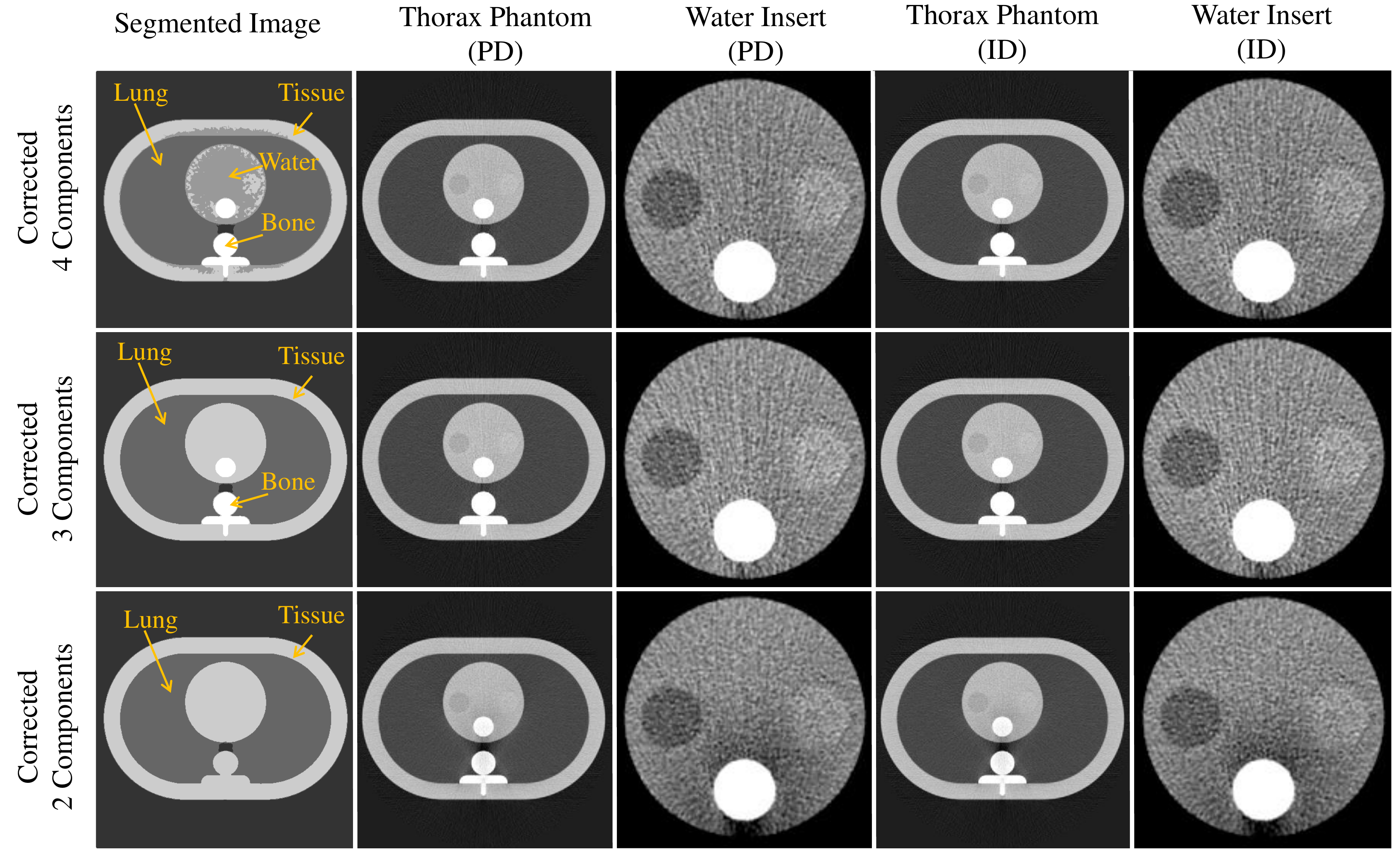}
    \vspace{-1em}
    \caption{The influence of the segmentation methods on the accuracy of the scatter correction is shown for the thorax phantom with water insert in both projection domain (PD) and image domain (ID). (First row) Results for the four components segmentation with lung, water, tissue and bone.
    (Second row) Results for the three components segmentation with lung, tissue and bone. (Third row) Results for the two components segmentation with only lung and tissue. Display window: [-1200HU, 500HU] for the thorax phantom images and [-300HU, 300HU] for the water insert images.}
    \label{fig:segmentaion}
\end{figure}

The influence of different segmentation methods on the accuracy of the scatter correction for the thorax phantom is depicted in Fig.~\ref{fig:segmentaion}. 

\vspace{-2mm}
\subsection{Experimental phantom and patient study}
\begin{figure}[t]
    \centering
    \includegraphics[width=3.5in]{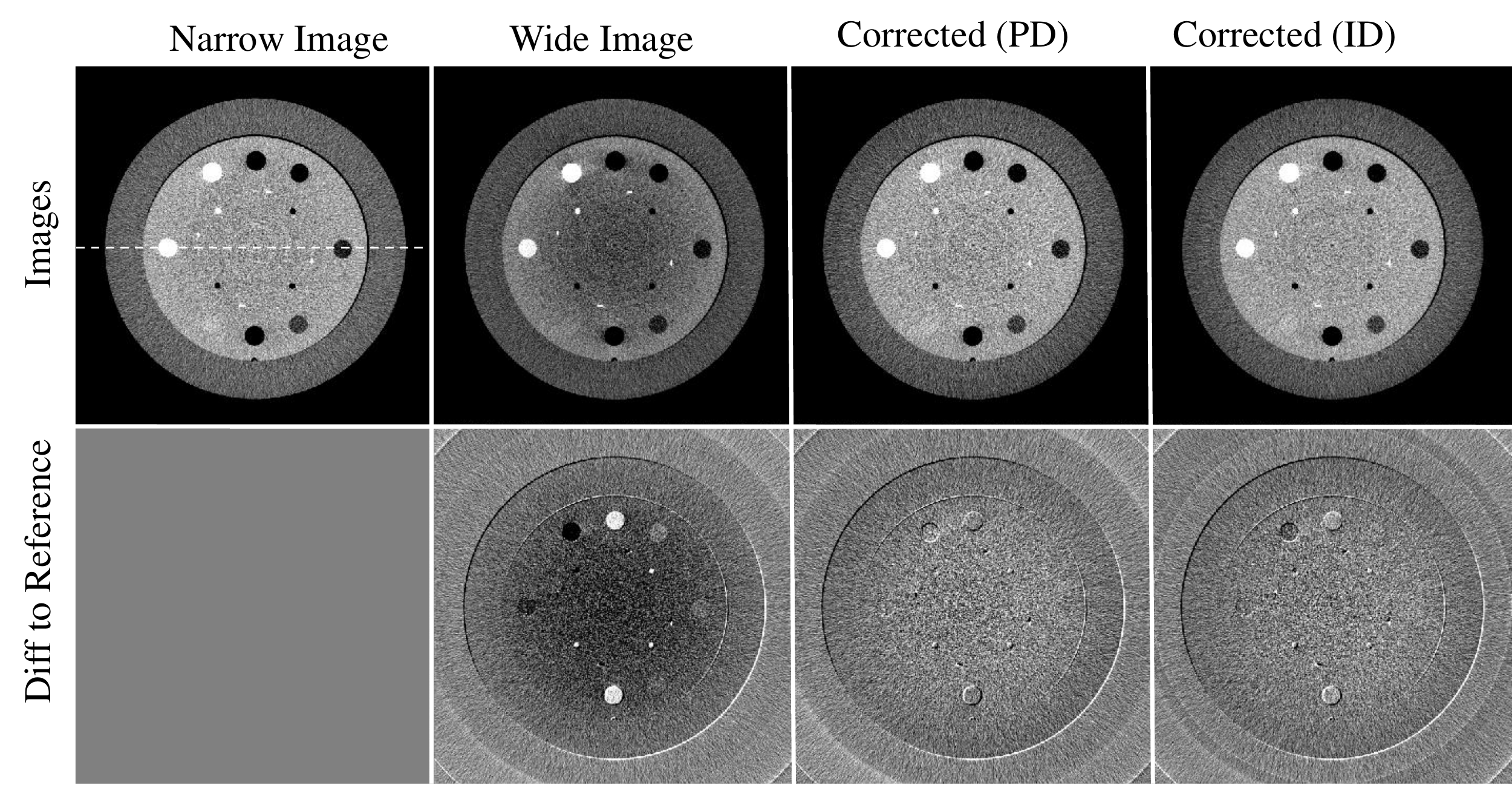}
    \caption{Catphan$^{\textregistered}$500 phantom with and without scatter correction. The narrow collimation image is considered to be the scatter-free image. Shading and reduced HU are shown in the wide collimation image.
    After correction, scatter artifacts are greatly reduced and HU are successfully recovered for both projection domain (PD) and image domain methods (ID). The difference images show each image subtracted with the narrow collimation image. Note that the wide collimation scan and the narrow collimation scan are two independent scans and the registration can not be perfect. Display window: [-150 HU, 150 HU] for both the CBCT images and the difference images.}
    \label{fig:catphan}
    \vspace{-3mm}
\end{figure}

\begin{figure}[t]
    \centering
    \includegraphics[width=3.5in]{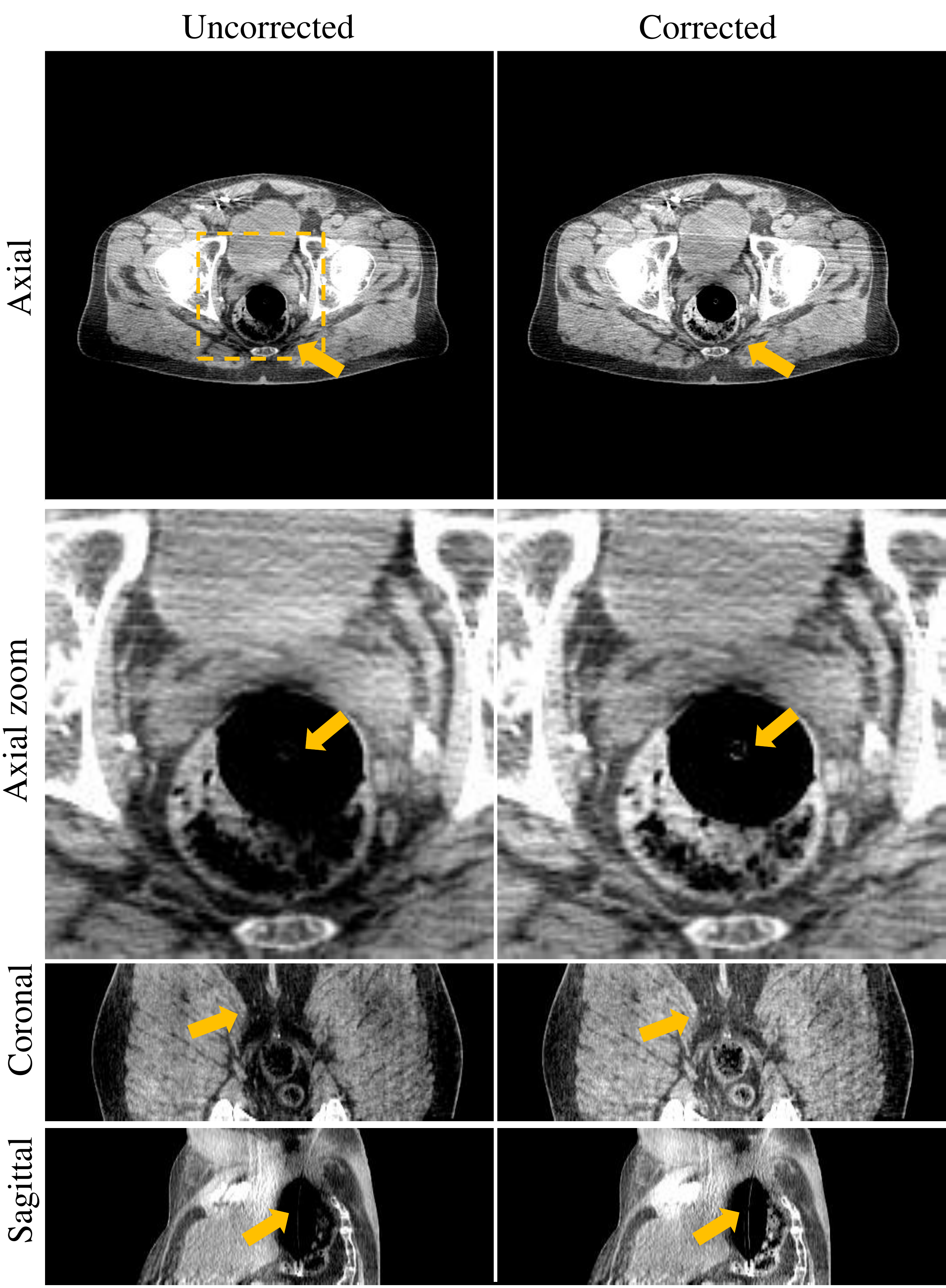}
    \caption{Scatter correction for a pelvis CBCT scan using the kV imaging system of the Varian TrueBeam system. Axial view, coronal view and sagittal view are depicted respectively. The uncorrected images acquired on the commercial system show shading artifacts and part of the anatomic structure was missing. After scatter correction, the shading artifacts were successfully reduced and the missing anatomic structure can be clearly seen. Display window: [-200 HU, 100 HU] for all images.}
    \label{fig:clinicalData}
    \vspace{-3mm}
\end{figure}
Fig.~\ref{fig:catphan} shows the CT images and the corresponding difference images of the Catphan$^{\textregistered}$500 phantom with and without scatter correction.
Fig.~\ref{fig:clinicalData} shows CT images of an \emph{in vivo} pelvis scan with the kilovoltage CBCT imaging system of the Varian TrueBeam system (Varian Medical Systems, Palo Alto, USA) without and with scatter correction.
After scatter correction, the HU number of the dark region was successfully recovered and the missing anatomical structures are recovered. It took about 1 min to correct a typical Varian clinical dataset ($512\times512\times81$) using a NVIDIA GeForce GTX~480 card.

\section{Discussion and conclusions}
\label{sec:discussion}

It has to note that the selection of the denoising parameters depends on segmented image. Since the noise level of the MC simulation data is  very high, a relatively large $K$ and $\beta$ (compared to the patient data) were used to smooth the segmentation error and to yield a denoised scatter that fit the true scatter well enough. To our belief, most of the clinical images have much lower noise level and the segmented images would be superior to the segmented images of the MC data in this study. Nevertheless, a larger $K$ and $\beta$ are recommended for badly segmented images to reinforce regularization.

In summary, this work presents a novel strategy to estimate a coarse scattered radiation profile, based on which a denoising algorithm was applied to yield the final scatter estimation. 
we demonstrated that a significant increase in image uniformity and HU accuracy were achieved after correction. The correction algorithm requires minimum computational cost, requires no modifications to existing system hardware, and can also handle cross-scatter for dual source dual energy CT scanner. Furthermore, the method can also be applied to spiral cone-beam CT where the anti-scatter grid can be uninstalled to further reduce radiation dose. 
\vspace{-2mm}
\section*{Acknowledgment}
The authors would like to thank Drs Jennifer Smilowitz, Guanghong Chen, Ke Li and Stephen~Brunner for providing the experimental data. The authors are also grateful to Dr Lei Xing for his careful reading of the manuscript and his useful comments.

\ifCLASSOPTIONcaptionsoff
  \newpage
\fi

\vspace{-2mm}


\bibliographystyle{IEEEtran}
%

%




\end{document}